\documentclass[%
 reprint,
 superscriptaddress,
 amsmath,amssymb,
 aps,
 prx,
]{revtex4-2}

\usepackage{graphicx}% Include figure files
\usepackage{dcolumn}% Align table columns on decimal point
\usepackage{bm}% bold math
\usepackage{hyperref}% add hypertext capabilities
\hypersetup{%
 setpagesize=false,%
 bookmarks=true,%
 bookmarksdepth=tocdepth,%
 bookmarksnumbered=true,%
 colorlinks=true,%
 breaklinks=true,   % splits links across lines
 pdfusetitle=true,  % \title and \author values into pdf metadata
 pdfsubject={},%
 pdfauthor={},%
 pdfkeywords={}}

\usepackage{physics}
\usepackage{siunitx}
\usepackage{color}
\usepackage{ulem}

\usepackage{verbatim}
\usepackage{siunitx}

\begin{document}

\preprint{APS/123-QED}

\title{Efficient low-energy single-electron detection using a large-area superconducting microstrip}% Force line breaks with \\

\author{Masato Shigefuji}
\email[]{m-shigefuji@g.ecc.u-tokyo.ac.jp}
%\thanks{}
\affiliation{Komaba Institute for Science (KIS), The University of Tokyo, Meguro, Tokyo 153-8902, Japan}

\author{Alto Osada}
\affiliation{Komaba Institute for Science (KIS), The University of Tokyo, Meguro, Tokyo 153-8902, Japan}
\affiliation{PRESTO, Japan Science and Technology Agency, Kawaguchi, Saitama 332-0012, Japan}

\author{Masahiro Yabuno}
\affiliation{Advanced ICT Research Institute, National Institute of Information and
Communications Technology (NICT), Kobe 651-2492, Japan}

\author{Shigehito Miki}
\affiliation{Advanced ICT Research Institute, National Institute of Information and
Communications Technology (NICT), Kobe 651-2492, Japan}
\affiliation{Graduate School of Engineering, Kobe University,
 Kobe 657-0013, Japan}

\author{Hirotaka Terai}
\affiliation{Advanced ICT Research Institute, National Institute of Information and
Communications Technology (NICT), Kobe 651-2492, Japan}

\author{Atsushi Noguchi}
\email[]{u-atsushi@g.ecc.u-tokyo.ac.jp}
\affiliation{Komaba Institute for Science (KIS), The University of Tokyo, Meguro, Tokyo 153-8902, Japan}%Lines break automatically or can be forced with \\
\affiliation{RIKEN Center for Quantum Computing (RQC), RIKEN, Wako, Saitama 351-0198, Japan}
\affiliation{Inamori Research Institute for Science (InaRIS), Kyoto, Kyoto 600-8411, Japan}


% Don't count these!
%TC:ignore
\begin{abstract}
Superconducting strip single-photon detectors (SSPDs) are excellent tools not only for single-photon detection but also for single-particle detection owing to their high detection efficiency, low dark counts, and low time jitter.
Although the detection of various particles, including electrons with \si{\kilo\electronvolt}-scale energy, has been reported so far, there have been no studies for detecting low-energy electrons.
It has yet to be clarified how low-energy electrons interact with electrons and/or phonons in a superconductor during electron detection.
Here we report the detection property of a superconducting micro-strip single-electron detector (SSED) for electrons with energy below 200~\si{\electronvolt}.
The detection efficiency is estimated as at least 37~\% when electrons impinging on the stripline possess an energy of 200~\si{\electronvolt}.
We also show that the minimum detectable energy of electrons is about 10~\si{\electronvolt} with our SSED, much lower than those of ions, which implies that the electron-electron interaction plays a significant role.
SSEDs might open a wide range of applications, from condensed matter physics to quantum information science, because of their compatibility with the cryogenic environment.
\end{abstract}
%TC:endignore

\maketitle

%\tableofcontents

%\section{\label{sec:level1}First-level heading:\protect\\ The line break was forced \lowercase{via} \textbackslash\textbackslash}
%\subsection{\label{sec:level2}Second-level heading: Formatting}
%\subsubsection{Wide text (A level-3 head)}
Among various superconducting detectors such as transition edge sensors~\cite{TES}, superconducting tunnel junction detectors~\cite{STJ}, and kinetic inductance detectors~\cite{KID}, superconducting strip single-photon detectors (SSPDs)~\cite{firstSSPD,SSPDreview,SSPDreview2,SSPDreview3,SSPDreview4} have attracted a great deal of attention owing to their high detection efficiency, low dark counts, and low time jitter.
SSPDs with high detection efficiency are already commercially available and are applied to a variety of fields, for example, quantum optics~\cite{SSPD_QuantumOptics}, quantum communication~\cite{SSPD_QKD,SSPD_QuantumCommunication}, optical telecommunication in the cosmic space~\cite{SSPD_OpticalTelecomCosmic}, light detection and ranging~\cite{SSPD_LIDAR}, and life science~\cite{SSPD_LifeScience}.

One interesting application of SSPDs is single-particle detection~\cite{SSPDparticle_review,SSPDparticle_review2}.
Various particles have been detected using SSPDs, for example, biopolymer ions~\cite{biopolymer1}, neutral molecules~\cite{neutral_molecules}, noble-gas ions~\cite{argon,helium}, and $\mathrm{\alpha}$ and $\mathrm{\beta}$ particles~\cite{alpha_and_beta_particles}.
In particular, an SSPD is utilized as a superconducting single-electron detector (SSED)~\cite{SSED} with near-unity efficiency for electrons with \si{\kilo\electronvolt}-scale energy.
SSPDs are excellent candidates for single-particle detectors for their fast response time, moderate operating temperature, and ability to detect massive molecules over 100~\si{\kilo\dalton}~\cite{SSPD_thicker,SSPDparticle_review}.

The detection mechanism of SSPDs relies on the excitation of quasiparticles in a superconducting strip-line by the energy transfer from an incident photon or particle.
However, there is a crucial difference between photon detection and particle detection.
Photons are absorbed by a superconductor and excite quasiparticles directly.
On the other hand, particles except for electrons are deposited on the surface of a superconductor and create many phonons, which excite quasiparticles through the electron-phonon interaction~\cite{SSPDparticle_review}.
This difference raises the question of how low-energy particles can be detected using SSPDs.
Neutral massive molecules with a quite-low kinetic energy of $\sim$0.1~\si{\electronvolt}, although the internal energy of molecules is not concerned, are detected using a 10~\si{\nano\metre}--12~\si{\nano\metre}-wide superconducting strip-line~\cite{neutral_molecules}.
On the contrary, an 800~\si{\nano\metre}-wide SSPD requires the kinetic energy of at least several hundreds of electronvolts to detect ions of noble gases such as argon~\cite{argon} and helium~\cite{helium}.
The experiment for the latter ions implies that over 99~\si{\%} of the kinetic energy of incident helium ions is not transferred to the electron subsystem~\cite{helium}.

\begin{figure}[b]
\includegraphics[clip, width=8.6cm]{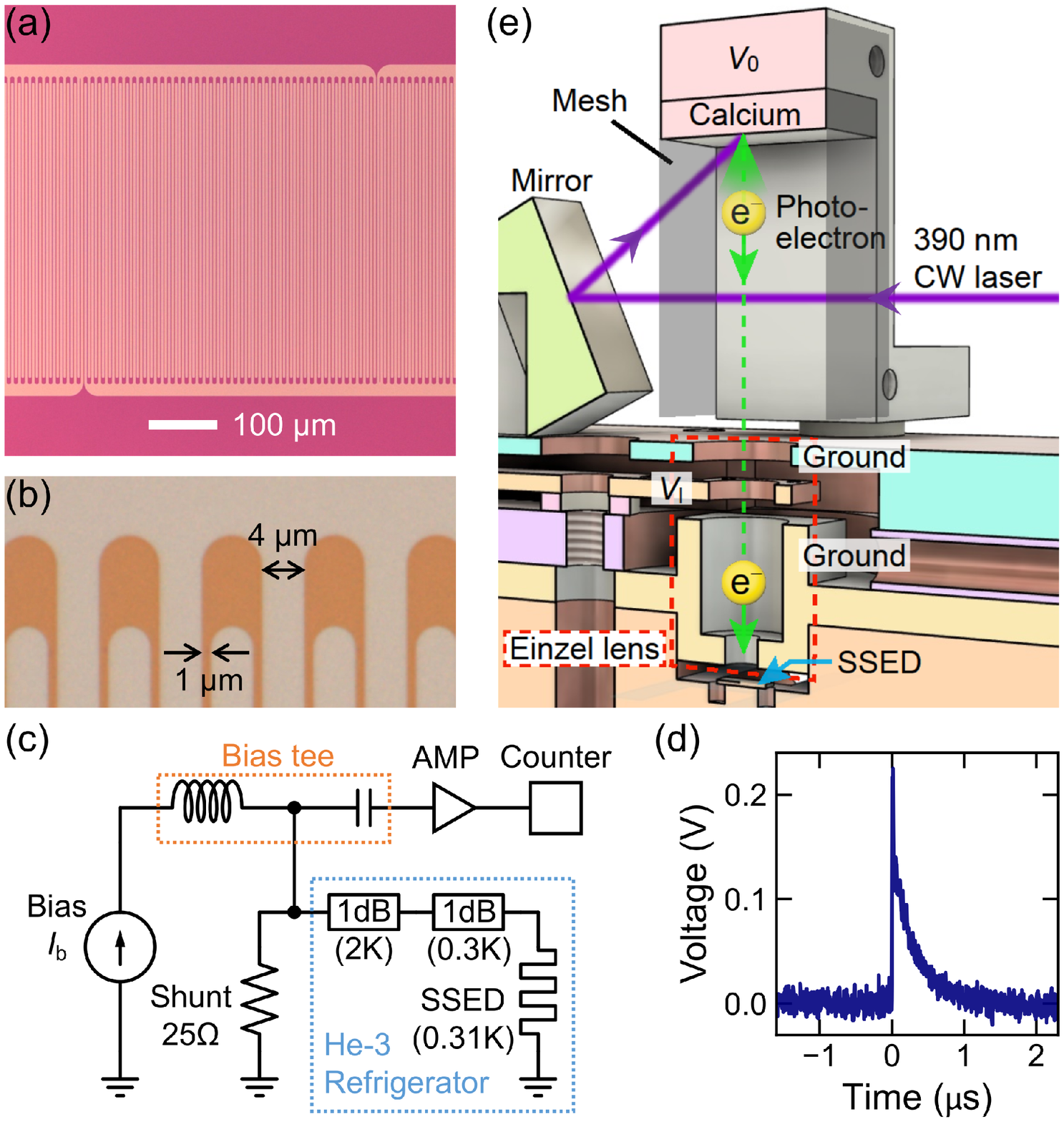}
\caption{\label{fig:1} (Color online) (a) Optical micrograph of the SSED with a large active area of ${420\times420}~\si{\square\micro\metre}$ with a filling factor of 20~\si{\%}. 
The NbTiN film is shown in pink. 
(b) Enlarged view of (a). 
The superconducting strip-line (orange) is 1~\si{\micro\metre} wide and 4~\si{\micro\metre} pitch. 
(c) Circuit diagram for electron detection. 
The SSED and two attenuators used as thermal anchors are put in a helium-3 refrigerator (enclosed by the dotted blue line).
(d) A typical detection signal from our SSED for $I_{\mathrm{b}}/I_{\mathrm{sw}}\approx0.95$, $V_{\mathrm{0}}=-100$~\si{\volt}, and $V_{\mathrm{l}}=0$~\si{\volt}.
(e) Schematic cross-sectional view of the experimental setup in the refrigerator. 
The 390~\si{\nano\metre}-wavelength CW laser is reflected by the mirror toward the calcium slab (solid purple line). 
Electrons are emitted by the photoelectric effect and accelerated toward the SSED (dashed yellow-green line).
The initial potential energy of electrons is controlled by a voltage of $V_{0}$.
Electrons are focused on the SSED using an einzel lens (enclosed by the dashed red line) with a voltage of $V_{\mathrm{l}}$.
}
\end{figure}

The detection of electrons is different from that of other particles mentioned above in the following two points: (i) they interact strongly with both electrons and phonons in a superconductor, and (ii) they can penetrate SSPDs.
Rosticher et al. (Refs.~\onlinecite{SSED}) have demonstrated, using an SSED, the detection of electrons with energies larger than 5 \si{\kilo\electronvolt}.
However, a natural question about how low-energy electrons an SSED can detect remains to be explored.

The dynamics of electrons with energy under several tens of electronvolts in solids is essential for measurement techniques such as low-energy electron diffraction (LEED) and photoemission spectroscopy.
Despite its importance, the mean free path of low-energy electrons in solids remains unclear due to a small number of experiments~\cite{IMFP_dev}.
Thus, exploring low-energy single-electron detection may give us valuable insight into the particle-detection application of SSPDs and the inelastic dynamics of electrons in solids.
Furthermore, such electron detectors could lead to various applications ranging from condensed matter physics, such as quantum electron microscopy~\cite{QEM_SSqubit,QEM} and cryogenic LEED~\cite{LEED}, to quantum information science, for example, an on-chip detector of electrons in a Paul trap~\cite{haffner2021,haffner2022,hybrid_quantum_system,osada2022}.

In this letter, we report the detection of electrons with kinetic energies ranging from $\sim$10~\si{\electronvolt} to 200~\si{\electronvolt} using a micro-strip SSED at a temperature of 310~\si{\milli\kelvin}.
We measure the detection efficiency for various electron energies, from which the minimum detectable energy is estimated. 
The system detection efficiency is measured as a function of a bias current through the micro-strip, and the detection mechanism of electrons is discussed.

Fig.~\ref{fig:1}(a) and (b) show optical micrographs of an SSED used in this study. 
The detector is fabricated by sputtering a $\approx$6~\si{\nano\metre}-thick NbTiN film on a silicon wafer with a 260~\si{\nano\metre}-thick $\mathrm{SiO_2}$ surface layer. 
We adopt our SSED with a \si{\micro\metre}-scale width structure to make its active area large~\cite{microstrip,large_area}.
The superconducting meandering micro-strip is 36~$\si{\milli\metre}$ long, 1~$\si{\micro\metre}$ wide, and 4~$\si{\micro\metre}$ pitch with an active area of $420\times420$~\si{\square\micro\metre} and a filling factor of $f=20~\si{\%}$. 
Fig.~\ref{fig:1}(c) displays the circuit diagram.
The SSED, put in a helium-3 refrigerator at a temperature of 310~$\si{\milli\kelvin}$, is shunted by a resistance of 25~\si{\ohm}.
A bias current $I_{\mathrm{b}}$ is applied to the SSED through a bias tee, which yields the measured switching current of $I_{\mathrm{sw}}\approx134~\si{\micro\ampere}$. 
Fig.~\ref{fig:1}(d) shows the typical detection signal from our SSED, amplified by the voltage gain of 125.
The decay constant is about 280~\si{\nano\second}.
There seems to be no difference between photon and electron signals.

\begin{figure}[t]
\includegraphics[clip, width=8cm]{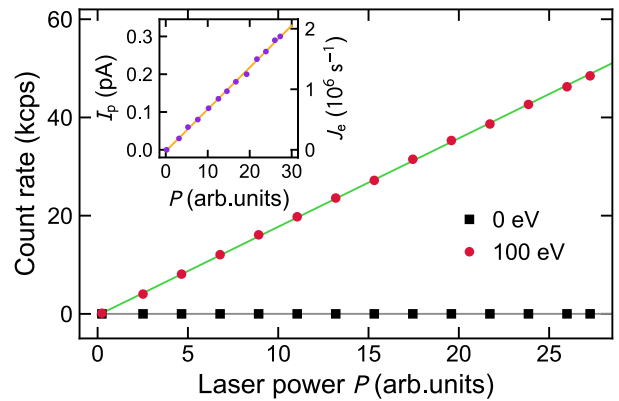}
\caption{\label{fig:2} (Color online) Count rate as a function of the 390~\si{\nano\metre}-wavelength laser power $P$ for $V_{\mathrm{0}}=0$~\si{\volt} (black squares) and $V_{\mathrm{0}}=-100$~\si{\volt} (red circles). 
A bias current of $I_{\mathrm{b}}/I_{\mathrm{sw}}\approx0.95$ is applied to the SSED. 
%The standard deviation of the count rate is negligibly small. 
(Inset) Photocurrent $I_{\mathrm{p}}$ flowing through the calcium slab and corresponding electron flux $J_{\mathrm{e}}$ as a function of $P$. 
%The electron flux $J_{\mathrm{e}}$ is calculated by dividing the current by the elementary charge. 
Solid lines (green, gray, and orange) are linear fit with an offset.}
\end{figure}

Fig.~\ref{fig:1}(e) schematically depicts the evaluation system of our SSED. 
Electrons are emitted by the photoelectric effect from a calcium slab illuminated by a 390~$\si{\nano\metre}$-wavelength CW laser. 
A surface oxide layer on the calcium slab, which disturbs the photoelectric effect, is eliminated by laser ablation using a nanosecond pulsed YAG laser. 
A photocurrent flowing between the calcium slab and ground is measured, which allows us to estimate the electron flux emitted from the slab. 
Electrons are accelerated toward the SSED and detected as voltage pulses through the ac line of the bias tee. 
A voltage $V_{\mathrm{0}}$ is applied to the calcium slab to modify the kinetic energy of incident electrons calculated as $E_{\mathrm{k}}=-eV_{\mathrm{0}}$, where $e(>0)$ is the elementary charge. 
Here, we neglect the initial kinetic energy $\lesssim$0.3~\si{\electronvolt} of electrons acquired in the photoelectric effect. 
Trajectories of electrons are modified through an einzel lens, which is consisted of a cylindrical hole with a voltage $V_{\mathrm{l}}$ sandwiched between grounded ones. 
Notably, the einzel lens affects only the electron trajectories, not the kinetic energy of incident electrons. 
The traveling distance of an electron is $\sim$30~$\si{\milli\metre}$.

%\section{Results}

The inset graph of Fig.~\ref{fig:2} shows that a current $I_{\mathrm{p}}$ flowing between the calcium slab and ground increases linearly with the laser power, which indicates the occurrence of the photoelectric effect.
The electron flux, given by dividing the current by the elementary charge, is converted from the laser power owing to the linear relationship between them.

\begin{figure}[t]
\includegraphics[clip, width=8.6cm]{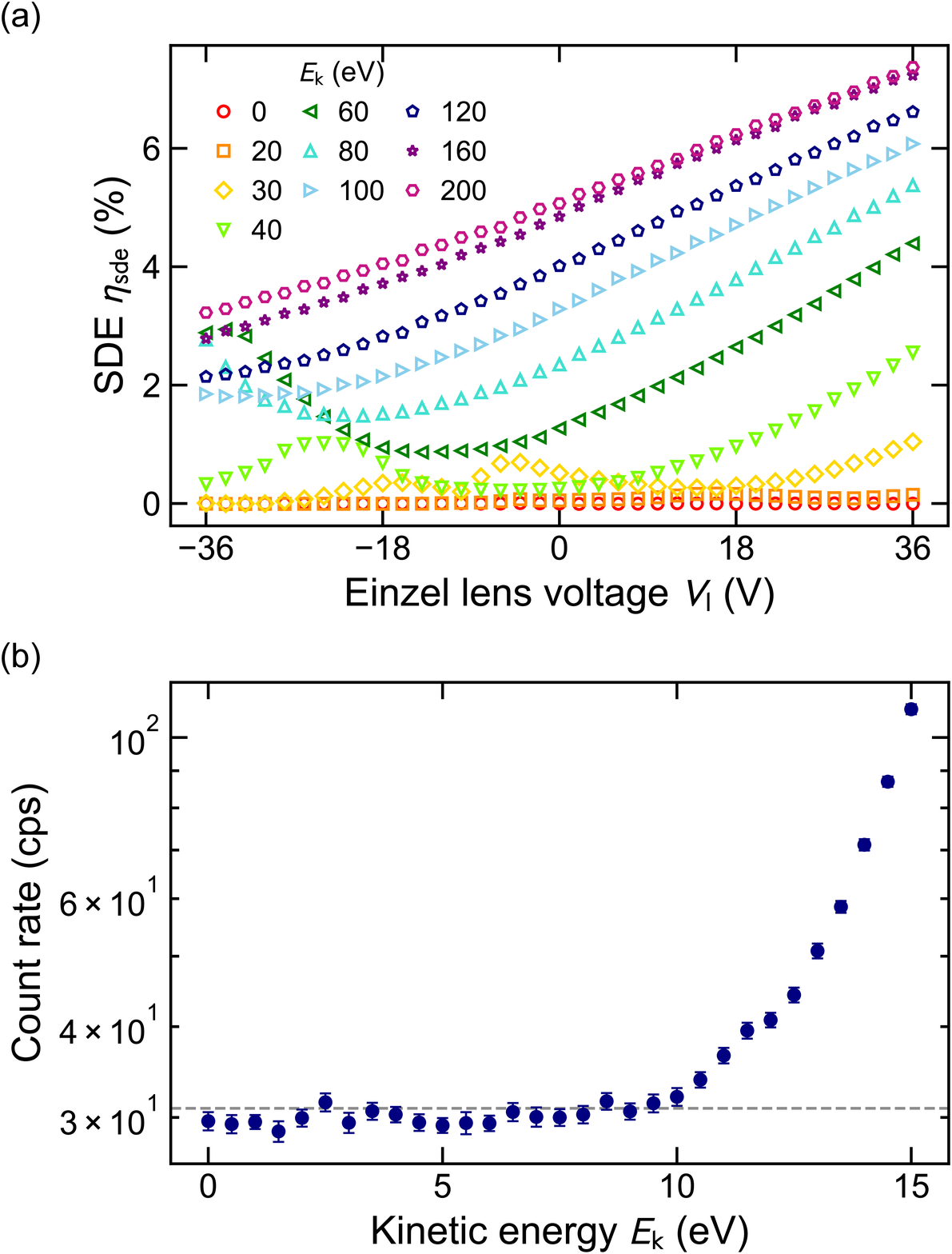}
\caption{\label{fig:3} (Color online) (a) System detection efficiency (SDE) $\eta_{\mathrm{sde}}$ as a function of a voltage applied to the einzel lens $V_{\mathrm{l}}$ for $I_{\mathrm{b}}/I_{\mathrm{sw}}\approx0.99$.
Correspondence between markers and electron energies $E_{\mathrm{k}}$ is shown in the panel.
(b) Count rate as a function of $E_{\mathrm{k}}$ for $V_{\mathrm{l}}=16$~\si{\volt} and $I_{\mathrm{b}}/I_{\mathrm{sw}}\approx0.98$. 
The dashed line stands for the dark count rate, namely, the count rate for $V_{\mathrm{0}}=V_{\mathrm{l}}=0$~\si{\volt}.
The error bars represent one standard deviation.
}
\end{figure}

This data convince us that the Coulomb interaction between electrons can be neglected for the following reason.
It takes tens of nanoseconds for an electron of $E_{\mathrm{k}}\sim10$~\si{\electronvolt} to arrive at the SSED from the calcium slab.
In contrast, the average emission interval of electrons is on the order of several microseconds calculated from the measured photocurrent.
Thus, the probability of more than one electron in the trajectory is very low, and the Coulomb repulsion between electrons and the resulting influence on their trajectories can be neglected.

To demonstrate that our SSED can detect low-energy electrons, we measure the count rate as a function of the power of 390~\si{\nano\metre}-wavelength laser for two values of $V_0$. 
The bias current and the voltage of the einzel lens are set to be $I_{\mathrm{b}}/I_{\mathrm{sw}}\approx0.95$ and $V_{\mathrm{l}}=0$~\si{\volt}, respectively. 
The result is shown in Fig.~\ref{fig:2}. 
The count rate for $V_{\mathrm{0}}=-100$~\si{\volt} (red circles) increases more rapidly with respect to the light power than that for $V_{\mathrm{0}}=0$~\si{\volt} (black squares). 
Since the detection of electrons, unlike photons, should be affected by the voltage $V_{\mathrm{0}}$, or the kinetic energy of electrons, the above behavior signals that our SSED has detected electrons.
The count rate for $V_{\mathrm{0}}=0$~\si{\volt} is slightly proportional to laser power and corresponds to the dark count rate (DCR), including the photon count of the stray 390~\si{\nano\metre}-wavelength light.
We convince that this rate does not include signals of electrons because it equals the count rate for $V_{\mathrm{0}}>0$~\si{\volt} where electrons cannot arrive at the SSED.

Fig.~\ref{fig:3}(a) plots the system detection efficiency (SDE) as a function of a voltage applied to the einzel lens $V_{\mathrm{l}}$ for several kinetic energies $E_{\mathrm{k}}$. 
Here we define the SDE as $\eta_{\mathrm{sde}}=[R(P,I_{\mathrm{b}},V_0,V_{\mathrm{l}})-R(P,I_{\mathrm{b}},0,0)]/J_{\mathrm{e}}(P)$, where $R$ and $J_{\mathrm{e}}$ are the count rate and the electron flux, respectively. 
The SDE can be alternatively decomposed as
\begin{equation}
    \eta_{\mathrm{sde}}=\eta_{\mathrm{cpl}}\eta_{\mathrm{abs}}\eta_{\mathrm{ide}},\label{eq:efficiency}
\end{equation}
where $\eta_{\mathrm{cpl}}$ is the coupling efficiency that photoelectrons arrive at the active area of the SSED, $\eta_{\mathrm{abs}}$ the absorption efficiency, and $\eta_{\mathrm{ide}}$ the internal detection efficiency which is an efficiency of generating a detection pulse after absorption of an electron~\cite{SSPDreview}. 
The einzel lens, which focuses electrons on the SSED, affects $\eta_{\mathrm{cpl}}$ and thus the SDE as shown in peaks in Fig.~\ref{fig:3}(a).

We identify the lower limit of the detection efficiency when electrons impinge on the superconducting region of the SSED as $\eta_{\mathrm{sde}}/f$, assuming that $\eta_{\mathrm{cpl}}=1$ and electrons are detected only in the superconducting region.
Here $f=0.2$ is the filling factor of the SSED.
For electron energy of $E_{\mathrm{k}}=200$~\si{\electronvolt}, the lower limit of the efficiency is estimated as about 37~\si{\%} ($V_{\mathrm{l}}=36$~\si{\volt}). 
It should be noted, however, that this value might be slightly overestimated because the previous work shows that even the non-superconducting region of an SSED can detect electrons with \si{\kilo\electronvolt}-scale energy~\cite{SSED}.

\begin{figure}[t]
\includegraphics[clip, width=8.6cm]{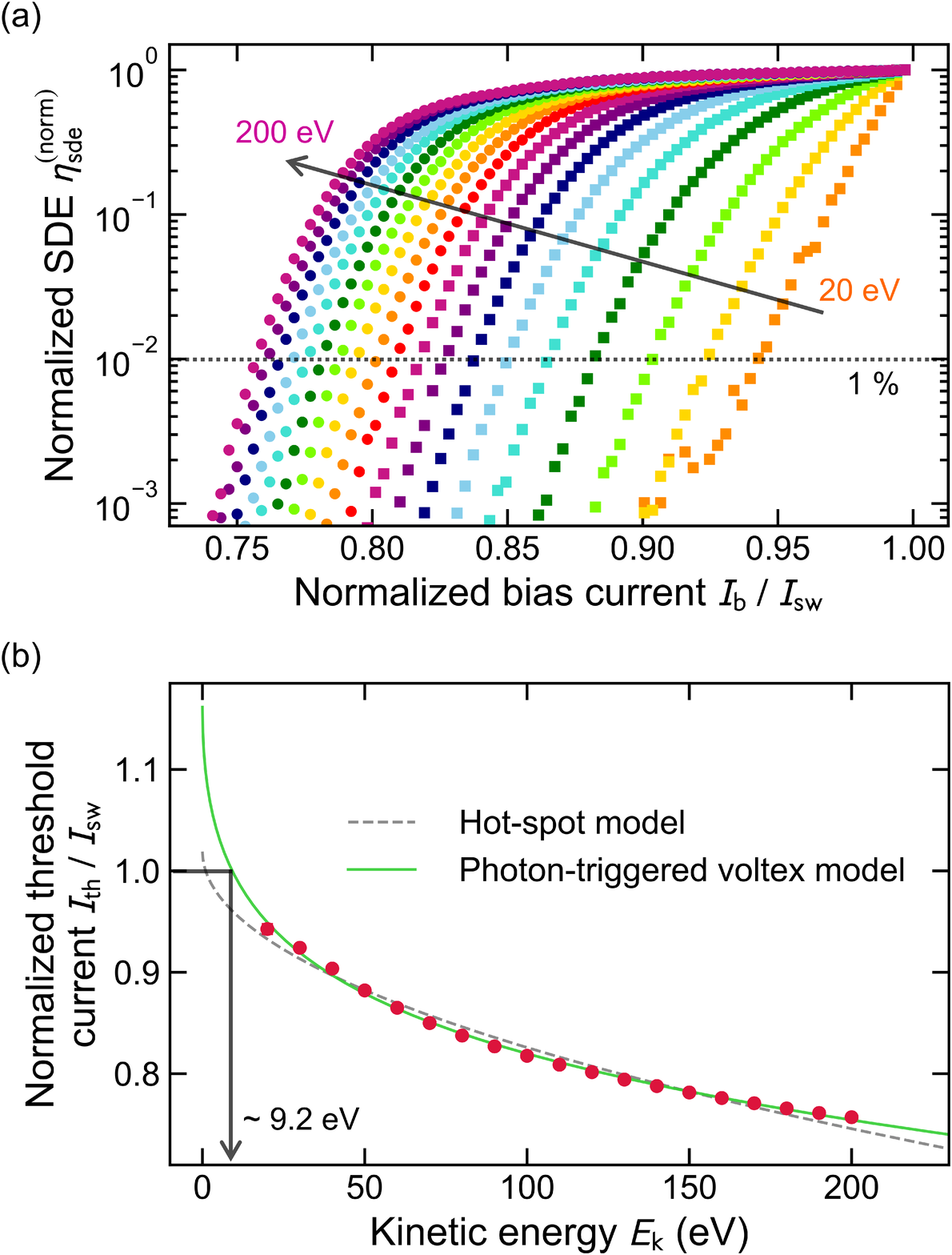}
\caption{\label{fig:4} (Color online) (a) Normalized system detection efficiency (SDE) $\eta_{\mathrm{sde}}^{\mathrm{(norm)}}=\eta_{\mathrm{sde}}/\eta_{\mathrm{sde}}^{\mathrm{(max)}}$ as a function of a normalized bias current $I_{\mathrm{b}}/I_{\mathrm{sw}}$ measured at dozens of kinetic energies ${E_{\mathrm{k}}}$.
${E_{\mathrm{k}}}$ increases by 10~\si{\electronvolt} from 20~\si{\electronvolt} to 200~\si{\electronvolt} in the direction of the arrow.
The squares and the circles represent 20~\si{\electronvolt}--100~\si{\electronvolt} and 110~\si{\electronvolt}--200~\si{\electronvolt}, respectively.
$V_{\mathrm{l}}$ is set to be 36~\si{\volt}.
The intersection of a $\eta_{\mathrm{sde}}^{\mathrm{(norm)}}$ curve and the dotted line of $\eta_{\mathrm{sde}}^{\mathrm{(norm)}}=1$~\si{\%} defines a normalized threshold current.
(b) Normalized threshold current $I_{\mathrm{th}}/I_{\mathrm{sw}}$ as a function of ${E_{\mathrm{k}}}$.
The dashed gray and the solid green curve are fitting curves using the normal-core hot spot model and the photon-triggered vortex model, respectively. 
}
\end{figure}

To examine the minimum detectable energy of our SSED, we fix $V_{\mathrm{l}}=16$~\si{\volt} and measure the count rate by changing $E_{\mathrm{k}}$ from 0~\si{\electronvolt} to 15~\si{\electronvolt}. 
The result is shown in Fig.~\ref{fig:3}(b). 
The count rate stays near the DCR (dashed line) until $E_{\mathrm{k}}\sim10$~\si{\electronvolt}, while it increases monotonically in the range of $E_{\mathrm{k}}\gtrsim10$~\si{\electronvolt} owing to the increase of energy transferred from electrons to the SSED.
Thus, the minimum detectable energy in our experimental setup is around 10~\si{\electronvolt}.
This value is an order of magnitude lower than that of ions, implying that the electron-electron interaction enables electrons to give energy efficiently to the electron subsystem in a superconductor.

The natural question is what determines the minimum detectable energy.
We propose two possible reasons: (i) inefficient energy transfer due to the long inelastic mean free path or (ii) the possibility that the switching current is lower than the depairing critical current.
We will explain the latter after showing Fig.~\ref{fig:4}(b) and here discuss the former.
The inelastic mean free path (IMFP)~\cite{IMFP_review} $\lambda$, the average distance an electron travels through a solid before losing energy, is roughly described by a universal curve independent of materials.
The IMFP increases rapidly with decreasing electron energy in ${E_{\mathrm{k}}}\lesssim50$~\si{\electronvolt}~\cite{IMFP_low}.
Consequently, the lower the energy, the more difficult it is to give energy to the SSED, resulting in the minimum detectable energy of a few tens of electronvolt.

We have to increase $\eta_{\mathrm{cpl}}$, $\eta_{\mathrm{abs}}$ and/or $\eta_{\mathrm{ide}}$ to enhance the SDE [see Eqs.~(\ref{eq:efficiency})].
An SSED with a large filling factor is preferred to increase $\eta_{\mathrm{cpl}}$, while a narrower strip-line is preferable for increasing $\eta_{\mathrm{ide}}$~\cite{helium}.
A thicker strip-line may benefit the enhancement of $\eta_{\mathrm{abs}}$~\cite{SSPD_thicker} for low-energy electrons~\cite{IMFP_low}, although this would entail a reduction in $\eta_{\mathrm{ide}}$~\cite{thickness}.
Thus, the thickness would need to be optimized.
Charged surface contaminants, mainly electrons trapped on the surface of the nonconductive region of the SSED, might work as a potential barrier~\cite{SSED}, and the cleaning of the surface of the detector might be effective.

We measure the normalized SDE $\eta_{\mathrm{sde}}^{\mathrm{(norm)}}=\eta_{\mathrm{sde}}/\eta_{\mathrm{sde}}^{\mathrm{(max)}}$ as a function of a bias current measured at dozens of $E_{\mathrm{k}}$'s.
The result is shown in Fig.~\ref{fig:4}(a).
The normalized SDE increases with kinetic energy and/or a bias current.
Even for a large bias current, the SDE increases gradually and does not seem to saturate.
We hypothesize that this is because the energy transferred from electrons to the SSED varies from zero to $E_{\mathrm{k}}$, and a larger bias current enables the detection of lower-energy dissipation events.

To investigate the electron-detection mechanism of the SSED, we calculate the threshold current $I_{\mathrm{th}}$ which is defined as a current for $\eta_{\mathrm{sde}}^{\mathrm{(norm)}}=1$~\si{\%} [shown as a dotted line in Fig.~\ref{fig:4}(a)].
The dependence of the normalised threshold current $I_{\mathrm{th}}/I_{\mathrm{sw}}$ on $E_{\mathrm{k}}$ is shown in Fig.~\ref{fig:4}(b).
We fit the measured data using some of the existing detection models: the normal-core hot spot model~\cite{firstSSPD} (dashed gray curve) and the photon-triggered vortex model~\cite{photon-triggered_vortex} (solid green curve). 
The experimental data seem to fit the latter model well.
The value of $I_{\mathrm{th}}$ at $E_{\mathrm{k}}=0$~\si{\electronvolt}, which corresponds to the depairing critical current $I_{\mathrm{c}}^{(\mathrm{dep})}$, does not equal to $I_{\mathrm{sw}}$.
This difference indicates that $I_{\mathrm{sw}}$ is about 86~\% of $I_{\mathrm{c}}^{(\mathrm{dep})}$ due to such as the device geometry, possibly microscopic defects, and variations of the cross-sectional area of the strip~\cite{Engle_mechanism}.
Measurement of $I_{\mathrm{th}}$ as a function of $E_{\mathrm{k}}$ can be valuable for studying the inelastic scattering mechanisms of an electron in a solid. 

The interesting thing is that the fitting curve of the photon-triggered vortex model crosses the $I_{\mathrm{th}}/I_{\mathrm{sw}}=1$ line at $E_{\mathrm{k}}\approx9.2$~\si{\electronvolt}.
This is possibly responsible for the minimum detectable energy being $\sim$10~\si{\electronvolt} with our SSED.
If this interpretation is correct, there are two ways to lower the minimum detectable energy: (i) design an SSED so that $I_{\mathrm{sw}}$ is as close as possible to $I_{\mathrm{c}}^{(\mathrm{dep})}$, and (ii) sharpen the falling edge of the $I_{\mathrm{th}}$--$E_{\mathrm{k}}$ curve.
The former is accomplished by designing the device geometry and/or the fabrication process appropriately~\cite{SSPDreview3}, while to achieve the latter, an SSED should be narrower~\cite{photon-triggered_vortex}.

%\section{Conclusion}
In conclusion, we realized the detection of electrons with low kinetic energies ranging from $\sim$10~\si{\electronvolt} to 200~\si{\electronvolt} using a superconducting micro-strip single-electron detector (SSED).
Low-energy electrons are emitted by the photoelectric effect and focused on the SSED with an einzel lens.
We measure the count rate of low-energy electrons and extract the system detection efficiency (SDE) by calibrating the number of electrons by the photocurrent measurement.
We estimate the detection efficiency as at least 37~\% when an electron with 200~\si{\electronvolt} hits the superconducting strip-line.
With our experimental setup, the minimum detectable energy of electrons is about 10~\si{\electronvolt}, indicating a more efficient energy transfer from electrons to a superconductor than ions.
We measure the SDE as a function of the bias current and find that the detection mechanism of electrons seems to be explained by the photon-triggered vortex model.
This work provides an efficient tool for investigating low-energy electrons at cryogenic temperature, in the scope of quantum electron microscopy and low-energy electron diffraction to expand the frontier of material science, and trapped electrons as a novel carrier of quantum information.

% Don't count these!
%TC:ignore
\begin{acknowledgments}
% Put your acknowledgments here.
We acknowledge Ippei Nakamura, Kento Taniguchi, Genya Watanabe, Shotaro Shirai, and Yasunobu Nakamura for fruitful discussions.
MS acknowledges the WINGS-ABC program at the University of Tokyo.
This work is supported by JST ERATO MQM project (Grant No. JPMJER1601), JSPS KAKENHI (Grant No. 19H01821 and 22H01965), JST SPRING (Grant No. JPMJSP2108), and JST Moonshot R\&D (Grant No. JPMJMS2063 and JPMJMS2066).
\end{acknowledgments}
%TC:endignore

% Don't count these!
%TC:ignore
%\quickwordcount{main}
%\quickcharcount{main}
%\detailtexcount{main}
%TC:endignore

\bibliography{main}% Produces the bibliography via BibTeX.

\end{document}